# Relating IS Developers' Attitudes to Engagement

Sherlock A. Licorish and Stephen G. MacDonell
Department of Information Science
University of Otago
PO Box 56, Dunedin 9054, New Zealand
Email: sherlock.licorish@otago.ac.nz, stephen.macdonell@otago.ac.nz

**Abstract**

Increasing effort is being directed to understanding the personality profiles of highly engaged information systems (IS) developers and the impact of such profiles on development outcomes. However, there has been a lesser degree of attention paid to studying attitudes at a fine-grained level, and relating such attitudes to developers' in-process activities, in spite of the fact that social motivation theory notes the importance of such a relationship in general group work. We have therefore applied linguistic analysis, text mining and visualization, and statistical analysis techniques to artefacts developed by 474 developers to study these issues. Our results indicate that our sample of IS developers conveyed a range of attitudes while working to deliver systems features, and those practitioners who communicated the most were also the most engaged. Additionally, of eight linguistic dimensions considered, expressions regarding work and achievement, as well as insightful attitudes, were most closely related to developers' engagement. Accordingly, team diversity and the provision of active support for outcome-driven developers may contribute positively to maintaining team balance and performance.

**Keywords**: Attitudes, Engagement, Linguistic Analysis, Text Mining, Visualization

## INTRODUCTION

Concerns over information systems (IS) inadequacies, high project failure rates and the non-delivery of enduring benefits from IS projects have been ubiquitous and longstanding (Boehm 2006). In spite of numerous recommendations extolling the benefits of specific development methodologies and tools (Licorish et al. 2009) over many years, the likelihood of project success has not increased at a commensurate rate (Standish Group 2009). In light of this, there is a growing degree of acceptance that it is human factors such as communication issues and behavioural conflicts, rather than the use of any particular tool or method, that underscore the performance of information systems development (ISD) teams and contribute directly to the causes of stakeholder (dis)satisfaction (Abrahamsson et al. 2006). Accordingly, studying these topics should provide a suitable means for researchers to understand the human elements of the ISD process, and so make informed recommendations for process improvements.

For instance, studies have considered under which circumstances developers should work collectively given a need to coordinate and communicate around specific systems features (Bird et al. 2009), resulting in team composition recommendations. Similarly, our own previous work has provided insights into how different teams work given the nature of their portfolio of tasks (Licorish and MacDonell 2013b), and other work has addressed the role of practitioners' personalities in mediating involvement in team communication (Abrahamsson et al. 2006). This latter theme is particularly relevant to the research reported here, as while researchers have considered IS developers' personality profiles in a general way (Gorla and Lam 2004), prior research has not explored developers' attitudes in a fine-grained manner, particularly in relation to within-project engagement. We address this gap in this study, and apply psycholinguistics in conjunction with text mining and visualization techniques to examine diversity in IS developers' attitudes as expressed in their messages, and we examine whether these attitudes are related to developers' engagement during teamwork.

Our work makes several contributions. We extend prior work that has examined the way in which IS developers' attitudes vary and are related to their engagement during the ISD process. We use visualizations and statistical tests to reveal insights into the relationship between attitudes and engagement, and the understandings gained from our application of multiple techniques to study developers' artefacts may provide useful pointers for similar future work. Furthermore, we provide recommendations for individuals undertaking ISD project governance.

In the next section we present the study's background and motivation, and we outline our specific research direction. We then describe our research setting, introducing our measures in this section. We subsequently present our results, and thereafter, we discuss our findings. We outline the implications of our results and highlight directions for future research. Threats to our study are acknowledged, and we then draw conclusions.

## BACKGROUND AND MOTIVATION

According to well-established linguistic theories it is possible to discern attitudes within individuals' communications (Pennebaker and King 1999). Works examining language use have established that there are unique variations in individuals' linguistic styles, and so linguistic analysis of the content of textual communication can reveal much about those communicating (Pennebaker and King 1999; Pennebaker et al. 2003). Previous work has also successfully linked personality traits to text syntax (Gill and Oberlander 2002), and more fine-grained analyses examining individual linguistic dimensions to precisely assess attitudes have reported consistency between specific language use and individual attitudes. For instance, previous research has found elevated use of first-person plural pronouns (e.g., "we") during shared situations and among individuals that share close relationships, whereas relatively high use of self-references (e.g., "I") has been linked to individualistic attitudes (Pennebaker et al. 2003). This suggests that attitudes can be discerned regardless of the specific setting.

A number of prior studies have therefore examined IS developers' behaviours that are evident in their textual communication, with the goal of generating a better understanding of the precursors to, and consequences of, developers' behavioural processes. For instance, Rigby and Hassan (2007) employed text analysis approaches to identify Apache Open Source Software (OSS) developers' personality profiles from their mailing list exchanges. Their findings revealed that the top two communicators were less extroverted than the other project members, and they scored lower on the openness to experience personality trait than the general population of contributors to the project. In contrast, Bazelli et al.'s examination of the StackOverflow forum found popular contributors to be most extroverted (Bazelli et al. 2013). Other work has noted that top committers exhibited more openness to experience than less active members, but that all personality traits were represented during teamwork (Licorish and MacDonell 2014), implying that no specific personality configuration determines team success.

Other forms of enquiry have utilised questionnaire-based techniques to study developers' personalities and their influence. For instance, a study of 47 professional systems developers in ten Swedish software companies found significant associations between personality factors and developers' behaviours (Feldt et al. 2010). Gorla and Lam's (2004) study of the personalities of 92 high-performing IS professionals in Hong Kong uncovered that extroverted programmers outperformed those who were intuitive. Wang (2009) also offered support for linking personality to developers' performance when reviewing 116 IS project outputs; and the need for variations in personality among IS developers working in teams have been stressed by Trimmer et al. (2002).

Beyond personality, however, other studies have shown that the willingness of individuals to actively participate in knowledge sharing and to contribute to team performance is based on multiple factors, including social motivation, rewards and incentives, and cognitive factors (Bock and Kim 2002; Geen 1991). While there is some uncertainty regarding the effects of incentives and rewards on individuals' engagement (Bock and Kim 2002), social motivation theory has proved to be generally effective for predicting participation in teamwork (Geen 1991). According to social motivation theory, teams' inter-personal interactions and norms have an impact on individual members' motivation to perform (Geen 1991). Thus, certain behaviours may be associated with an individual's performance. For instance, those who express individualistic and negative sentiments (or operate in such norms) may be less motivated to contribute to team outputs, whereas those who are outcome-driven may be more committed to maintaining team performance. Similarly, as assessed by Trimmer et al.'s (2002) personality-based study, a group of individuals that *collectively* exhibits diverse attitudes and behaviours may be more effective at problem solving than a homogeneous group. In fact, role theories have indeed established that a degree of heterogeneity is necessary for positive team performance and maintaining team synergy (Belbin 2002). Accordingly, studying fine-grained measures of ISD team members' attitudes should provide insight into how attitudes are distributed during ISD, and the actual relationship (if any) between IS developers' specific attitudes and their engagement in ISD activities. These research avenues could inform team composition strategies, and provide an interesting platform for future work. We address this research opportunity by answering the following research questions:

RQ1. Do IS developers express diverse attitudes?

RQ2. Are IS developers' attitudes related to their engagement?

## RESEARCH SETTING

In order to address our research questions we examined development artefacts derived from a specific release (1.0.1) of Jazz (based on the IBM[R] Rational[R] Team Concert[TM] (RTC)[1]): a fully functional environment for developing software systems and for managing the ISD process. Jazz includes features for work planning and traceability, systems builds, code analysis, bug tracking and version control, among others. Changes to source code in the Jazz environment are permitted only as a consequence of a work item (WI) being created beforehand, such as a defect, a support task, or an enhancement request; and project communication, the primary content explored in this study, was enforced through the use of Jazz itself. The instance of Jazz studied comprised product and process data collected from distributed development and management activities across the USA, Canada and Europe. In Jazz each team includes multiple individual roles (e.g., programmer, admin, team lead), with a project manager responsible for overall leadership. Jazz teams use the Eclipse-way methodology for guiding the ISD process. This methodology outlines iteration cycles that are six to eight weeks in duration, generally conforming to agile principles. All information regarding the activities and outcomes of the development process is stored in a server repository accessible through an Eclipse-based (RTC) client interface.

---

[1] IBM, the IBM logo, ibm.com, and Rational are trademarks or registered trademarks of International Business Machines Corporation in the United States, other countries, or both.

We leveraged the Client API to extract team information and development and communication artifacts from the repository. In total we extracted 30,646 resolved WIs developed across 30 iterations between June 2005 and June 2008 by 474 developers. Practitioners communicated 117,101 messages around the 30,646 WIs. Given our intent to study the diversity of developers' attitudes and to relate these to their engagement, we used the practitioner as our unit of analysis, and employed multiple techniques to analyse their artefacts, as now discussed.

**Measuring Attitudes**: Given that language use is the key phenomenon under consideration in this work we study attitudes from a Big Five perspective. The Big Five personality model was developed from the theoretical stance that personality is encoded in natural language, and differences in personality are evident in linguistic variations (Goldberg 1990). As implied by its name, the Big Five personality model considers five personality traits, being extroversion, agreeableness, conscientiousness, emotional stability (or neuroticism) and openness to experience. Extroversion describes individuals' desire to seek company and their drive for stimulation from the external world. Agreeable individuals are said to be cooperative and sensitive to others. Conscientiousness denotes a preference for order and goal-directed work. Individuals who are emotionally unstable (or neurotic) have a tendency to show excessive negative emotions and anger. Finally, the openness to experience trait is associated with being insightful and open to new ideas. We used these dimensions as baselines for selecting four classes of attitudes that can be readily detected in language use, being social attitudes, cognitive attitudes, achievement attitudes and individualistic (negative) attitudes. To illustrate, social attitudes are assessed through the use of words such as "give", "beautiful" and "perfect", while words including "think", "consider" and "should" convey cognitive attitudes (Pennebaker et al. 2003). We selected these four forms of attitudes, as against the five noted in the Big Five model, because previous work had noted some level of overlap in linguistic usage across personality traits (e.g., positive emotion can be evident among extroverts and agreeable individuals (Pennebaker and King 1999)). We applied linguistic analysis, text mining and visualization procedures to our pre-processed data to reveal and understand developers' attitudes, as now introduced.

*Linguistic Analysis*: Drawing on previous work (Rigby and Hassan 2007) we employed the Linguistic Inquiry and Word Count (LIWC) software tool in our analysis of developers' attitudes. The LIWC was created after four decades of research using data collected across the USA, Canada and New Zealand (Pennebaker and King 1999). This tool captures over 86% of the words used during conversations. Written text is submitted as input to the tool in a file that is then processed and summarized based on the LIWC tool's dictionary. Each word in the file is searched for in the dictionary, and specific type counts are incremented based on the associated word category (if found), after which a percentage value is calculated by aggregating the number of words in each linguistic category over all words in the messages. For example, if there were 10 instances of words belonging to the "social" dimension in a message with a length of 200 words, then the percentage value for the "social" dimension would be (10/200=)5.0%. The different categories (or dimensions) in the LIWC output summary are said to capture the attitudes of individuals by assessing the words they use (Mairesse et al. 2007; Pennebaker and King 1999). Two distinct linguistic dimensions were selected to collectively model each of our four attitude classes, based on prior research. To study social attitudes we selected both social words (e.g., give, love) and positive words (e.g., beautiful, perfect), while long words (words with more than 6 letters) and insightful words (e.g., think, believe) were used to operationalize cognitive attitudes. Achievement attitudes were studied using work- (e.g., goal, delegate) and achievement-related terms (e.g., attain, resolve). Finally, individualistic attitudes were represented by self-focus (e.g., I, my) and negative words (e.g., hate, dislike). These categories and their associated words were taken directly from the previously validated LIWC dictionary.

*Text Mining and Visualization*: Given our intent to analyse multiple linguistic categories as representing developers' attitudes we sought an analysis approach capable of isolating these linguistic types while still considering the potential for relationships among them. Accordingly, we utilized the Self-organizing Map (SOM) unsupervised learning algorithm to reveal patterns in practitioners' attitudes as captured by the LIWC. Kohonen's SOM employs unsupervised algorithmic training to classify multidimensional data into similarity graphs and clusters (Kohonen 1998). This process groups similar vectors based on their relative *Euclidean* distance using a nonparametric, recursive regression process (Kohonen 1998). We visualized the output from the SOM clustering through the Viscovery SOMine package (viscovery.net). This package enabled us to model the clustered data onto a two-dimensional map (as illustrated in Figure 1), facilitating inspection and follow up statistical testing.

**Measuring Engagement**: Various approaches have been used to assess individual-level engagement in ISD tasks. Productivity-related measures such as lines of code (LOC) per unit of effort (Curtis 1981), time taken (Espinosa et al. 2007) and task changes completed (Cataldo and Herbsleb 2008) are among those that have been used. Along with others, Cataldo and Herbsleb (2008) argued that measures based on LOC are unreliable due to variability in developers' coding styles (e.g., some developers are more verbose). Time taken to complete development tasks may be confounded when there are feature inter-dependencies (e.g., a developer may start work on a feature that needs to use classes under development by another developer, and so may be delayed). We therefore used the number of task changes as indicative of developers' engagement. A developer was considered to change a task if they created, modified, or resolved that task; as per prior work (Cataldo and Herbsleb 2008). We also compared these results to the volume of messages communicated, as reported next.

## RESULTS

As noted above we extracted measures for linguistic classes representing social attitudes, cognitive attitudes, achievement attitudes and individualistic (negative) attitudes according to the LIWC corpus for the 117,101 messages in release 1.0.1 of Jazz. Given that the individual developer was our unit of analysis the relevant linguistic measures were aggregated into a vector to represent each of the 474 contributors. Not unexpectedly, message distribution was uneven: a few developers submitted as little as one message, while the maximum number of messages contributed by a single contributor was 5,403. We then allowed the SOM unsupervised learning algorithm to cluster practitioners based on their linguistic usage. The output from the Viscovery SOMine package is presented in the map in Figure 1, which shows that the 474 developers were automatically grouped into five clusters (C1 to C5) by the SOM algorithm. We examine each of these clusters and associated data in answering our two research questions in the subsections that follow.

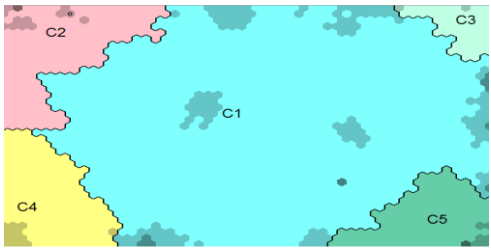

Figure 1. Overall cluster map for developers' linguistic usage

### Diversity of Attitudes (RQ1)

Figure 2 depicts how the eight linguistic dimensions were evident to varying degrees in the communications of the 474 practitioners, as visualized across eight component maps that reflect their relative use of social, positive, work focus, achievement focus, long, insightful, self-focus and negative words, respectively. For the *social* and *positive* dimensions (which together represented social attitudes in our analysis), some practitioners communicated with very few of these words (e.g., those represented in cluster C4 of the *social* and *positive* maps), whereas for other team members up to 50% of their discourses were social (e.g., see the red segments of cluster C1 in the *social* map) and up to 82% of some members' discourses were positive in nature (e.g., see the red segment of C5 in the *positive* map). Note also that these two sub-groups are depicted relatively closely on the maps, indicating a high degree of overall similarity between these two sub-groups of practitioners. *Work* and *achievement* language use was also concentrated in those represented towards the bottom of cluster C1 and towards cluster C4; and, although the maximum values for these dimensions were lower than those previously mentioned (40% and 29% respectively), achievement attitude was expressed in general by a large(r) spread of developers. This dispersion was also evident for what are considered to be *long* words in the LIWC, which were used by most of the Jazz practitioners. Within that group was a particular sub-group of practitioners (in C1) who utilized a greater proportion of *insightful* words. Figure 2 also shows that there was an overlap in the relative use of *self-focus* and *negative* words, and particularly for practitioners who were represented in the upper areas of cluster C1 and in cluster C3.

Notwithstanding scale variations, the component maps in Figure 2 demonstrate some informative patterns, and overlaps, regarding relative linguistic use. Table 1 presents the mean percentage usage of the linguistic dimensions for each cluster, which shows that cluster C1 comprised the highest number of practitioners (309 of 474), whose members typically conveyed a comparatively high degree of individualistic attitudes (3.8% and 5.4% respectively). Those members assigned to cluster C3 (19 of the 474) tended to express around twice the extent of individualistic attitude (5.4% and 7.3% respectively) as the overall Jazz average, they exhibited minimal achievement attitude (2.1% of work related utterances), and much lower levels of social attitude (3.5% and 3.9% respectively) than those in the other clusters, and the Jazz team average. In contrast, 59 of their colleagues clustered in C2 appeared to be twice as social as those in the other clusters (7.2% of social discourses) and exhibited little in the way of individualistic attitudes (2.6% and 1.3% respectively). While those 44 practitioners who were clustered in C4 were not highly social, these members were the most achievement-driven (3.9% and 3.5% respectively). Finally, the 43 members belonging to cluster C5 were most positive (communicating with 29% of positive utterances, and more than three times the Jazz team's average), but exhibited little cognitive (only 1.0% insightful utterances), achievement (1.8% and 1.2% respectively) and individualistic (1.2% and 0.9% respectively) attitudes.

We applied formal statistical testing to evaluate the significance of these results and to explore if developer usage of particular linguistic types was interrelated. We first used the Shapiro-Wilk test to check the normality of the various linguistic distributions. The results of these tests confirmed that the data for the eight linguistic distributions significantly deviated from normal ($p < 0.05$). We then performed non-parametric Kendall tau-b correlation tests, the results of which are provided in Table 2. In Table 2 it is notable that, although some of these results were statistically significant and positive, many of them were not strong. Of further note, however, is the evidence in Table 2 that practitioners who used positive language also used long words (a medium correlation result), and those who were work-focused also communicated with a significant amount of achievement language (this relationship was strong). This linkage between those who were work- and achievement-focused confirms the results drawn from the SOM analysis (refer to Figure 2 for visualizations).

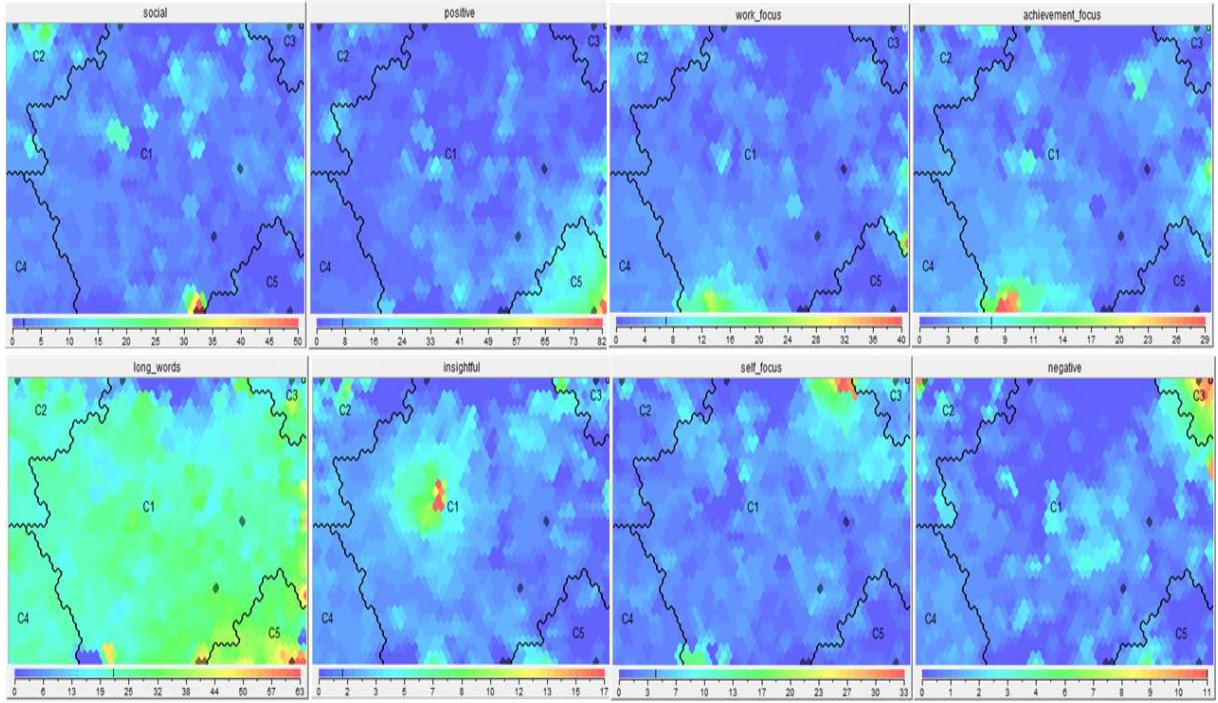

Figure 2. Component maps for developers' linguistic usage

Table 1. Cluster Measures for Linguistic Usage (Mean Percentage Usage)

| Linguistic Dimension | | Mean percentage usage per cluster (number of members) | | | | | Jazz mean |
|---|---|---|---|---|---|---|---|
| | | C1 (309) | C2 (59) | C3 (19) | C4 (44) | C5 (43) | |
| Social attitudes | social | 3.9 | 7.2 | 3.5 | 3.8 | 1.2 | 3.9 |
| | positive | 5.9 | 4.8 | 3.9 | 4.2 | 29.6 | 9.7 |
| Cognitive attitudes | long words | 22.7 | 19.3 | 21.3 | 22.0 | 34.1 | 23.9 |
| | insightful | 2.1 | 2.0 | 2.4 | 1.9 | 1.0 | 1.9 |
| Achievement attitudes | work-focus | 3.4 | 3.7 | 2.1 | 3.9 | 1.8 | 3.0 |
| | achievement-focus | 3.1 | 3.0 | 2.7 | 3.5 | 1.2 | 2.7 |
| Individualistic attitudes | self-focus | 3.8 | 2.6 | 5.4 | 2.5 | 1.2 | 3.1 |
| | negative | 5.4 | 1.3 | 7.3 | 1.2 | 0.9 | 3.2 |

**Attitudes and Engagement (RQ2)**

We also utilized SOMs to explore the extent to which developers shared messages while working on WIs and to examine how these members were involved in creating, modifying and resolving ISD tasks. These results are depicted in the component maps of Figure 3, which show that the members clustered in C4 dominated all four activities. In fact, the component maps for comment count, number of tasks created, number of modifications, and number of tasks resolved demonstrate that there was almost an identical overlap in values. We triangulated these results through formal statistical testing after first testing the data for normality, which confirmed skewness in the distributions. Given the overlap in values in Figure 3, we aggregated the three forms of development activities (numbers of tasks created, modified and resolved) into one measure, '*task changes*'. A Kendall tau-b correlation test was then conducted to determine the strength of any relationship between the number of messages communicated by practitioners' and their engagement in task changes. This result confirmed that there was a strong, positive and statistically significant correlation: $\tau = 0.763$, $p < 0.01$.

We performed additional statistical testing to explore the relationships between developers' linguistic profiles and their engagement in task changes. These results, provided in Table 3, demonstrate that, of the insightful, work-focus, achievement-focus and negative linguistic dimensions, which were all significantly related ($p < 0.01$) to task changes, the strongest correlations existed between work- and achievement-focus and practitioners' engagement. Practitioners were also more actively engaged when they communicated negative words. The social, positive, long words and self-focus dimensions were not correlated with practitioners' engagement.

We then built a model to examine whether the linguistic dimensions interacted to affect developers' engagement in task changes. First, we used Kendall tau-b correlation tests to examine whether variations in message length affected the observed measures for practitioners' use of the various linguistic dimensions, as this had been found in previous research to affect team interactions (Sethi et al. 2001). The only noteworthy outcome from these tests related to the insightful linguistic dimension, which confirmed a small positive correlation that was statistically significant ($\tau = 0.223$, $p < 0.01$). We also included a control factor for practitioner role, given previous evidence that had established that members' status impacts on their engagement (Kirchler and Davis 1986). Finally, given that the distributions were all skewed, we performed a natural log transformation on the variables, prior to performing a stepwise multiple regression.

While a statistically significant model emerged ($F_{3,470} = 21.544$, $p < 0.01$), the Adjusted R-squared value revealed that our model accounted for just 12% of the variance in task changes. In keeping with the visual impression that was evident across Figures 2 and 3, our results indicated that work- and achievement-focus and insightful linguistic utterances were all significantly related to developers' involvement in task changes (standardized beta coefficients = 0.180, 0.140 and 0.127 respectively, $p < 0.01$). The use of work-related terms was the strongest single predictor of the extent to which developers engaged in task changes.

Table 2. Kendall Tau-b Correlation (τ) Results for Relationships in Linguistic Usage

| Factor | 1 | 2 | 3 | 4 | 5 | 6 | 7 | 8 |
|---|---|---|---|---|---|---|---|---|
| 1 social | 1.0 | -0.16* | -0.16* | 0.17* | 0.12* | 0.14* | 0.06 | 0.06 |
| 2 positive | | 1.0 | **0.29*** | -0.10* | 0.06 | 0.02 | -0.10* | 0.01 |
| 3 long words | | | 1.0 | -0.06 | 0.07* | 0.01 | -0.13* | 0.05 |
| 4 insightful | | | | 1.0 | 0.11* | 0.16* | 0.16* | 0.10* |
| 5 work-focus | | | | | 1.0 | **0.53*** | -0.02 | 0.11* |
| 6 achievement-focus | | | | | | 1.0 | 0.07* | 0.13* |
| 7 self-focus | | | | | | | 1.0 | 0.11* |
| 8 negative | | | | | | | | 1.0 |

Note: *$p < 0.05$; bold values represent noteworthy results

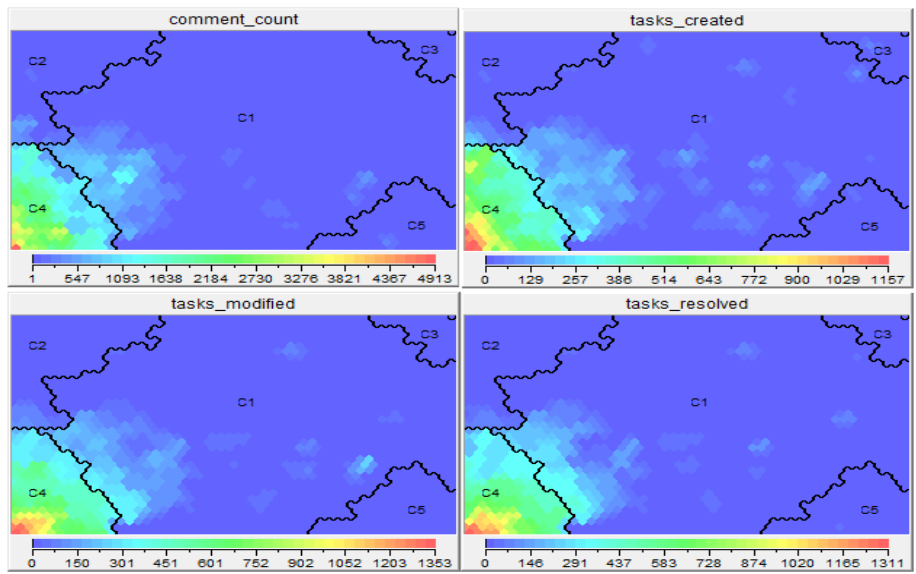

Figure 3. Component maps for developers' engagement

Table 3. Kendall Tau-b Correlation (τ) Results for Relationship Between Attitudes and Task Changes

| Linguistic Dimension | | Correlation Coefficient (τ) |
|---|---|---|
| Social attitudes | social | 0.04 |
| | positive | 0.06 |
| Cognitive attitudes | long words | 0.06 |
| | insightful | 0.01** |
| Achievement attitudes | work-focus | **0.22**** |
| | achievement-focus | **0.22**** |
| Individualistic attitudes | self-focus | -0.02 |
| | negative | 0.19** |

Note: **$p < 0.01$; bold values represent noteworthy results

## DISCUSSION

*Do IS developers express diverse attitudes?* Our results confirm that the 474 Jazz developers considered in this study collectively expressed heterogeneous attitudes when working to deliver systems features. While some developers were more social, others were more cognitive, others maintained high levels of achievement focus, and still others exhibited individualistic attitudes. Such a spread of attitudes is likely to have a balancing effect on the team's behavioural climate (Belbin 2002), a phenomenon that could be particularly beneficial in agile settings where interdependence is encouraged. For example, a social and positive outlook is generally necessary during times of high intensity and stressful teamwork, in terms of providing encouragement for developers faced with schedule and delivery pressure, whereas work-focused and outcome-driven attitudes are likely to be most effective during task analysis and brainstorming stages. These latter attitudes may also promote team urgency when developers are facing imminent deadlines, or where members are affected by outside distractions. Moderate levels of individualistic attitudes may also be useful for maintaining high team standards through critical and constructive debates. Of course, IS developers may also be more tolerant to their peers expressing such forms of attitudes during stressful project phases, or when individual focus is necessary.

Altogether however, this diversity in attitudes among developers may be beneficial for maintaining behavioural balance and enhancing team performance. These assertions are in fact supported by research conducted in management and role theories. For instance, Benne and Sheats (1948) observed three dimensions of individual behaviour in problem-solving teams: helpful and supportive behaviours (personal and social roles), task-concerned behaviours (task roles), and debate- and conflict-centred behaviours (individualistic roles). Benne and Sheats (1948) note that *all* roles are important during group tasks; that role requirements vary during different stages of teamwork; and that these roles are adopted by various individuals at different times, enabling group members to provide maximum team contributions and increase the likelihood of group success.

Apart from observing diversity in attitudes, our findings show that Jazz developers displayed some similarities across attitudes. For instance, those who were positive also used longer words during their exchanges. Similarly, those practitioners who were focused heavily on work tasks were also highly achievement driven. Personality theories have shown that socially inclined individuals generally seek the company of others, and are driven by stimulation from the external world (Goldberg 1981). Perhaps the practitioners who were so inclined in this work also exhibited more patience in articulating their thoughts, and hence, they were happy to spend time framing their discourses with longer words. On the other hand, conscientious individuals show a general preference for order and goal-directed work, and are achievement driven. Thus, the evidence for work and achievement focus is fitting in this context. The expression of such attitudes may be a useful indicator when assessing individuals' potential performance or commitment to team efforts. We further consider this issue next.

*Are IS developers' attitudes related to their engagement?* Our findings reveal that those developers who had the most to say were also most actively engaged in change log activities. We noted that those who said the most were not necessarily the most social and positive, a result that diverges somewhat from established thinking in the personality psychology space, which tends to associate talkers with the extroversion personality trait (Goldberg 1981). In fact, those applying personality theories to study the behaviours of ISD teams have also linked the need to communicate with some specific nominal roles, and particularly those that are dedicated to administering and coordinating (e.g., team leaders) (Andre et al. 2011). However, our evidence suggests that the size of the portfolio of tasks under a practitioner's consideration may influence their need to communicate and perform. This indicator may also be a stronger driver of a developer's level of engagement than their social orientation (as noted by an early study considering human factors during ISD (Curtis 1981)).

From an attitude perspective, our findings indicate that the expression of work, achievement and insightful attitudes were the strongest indicators of developers' engagement. It should be acknowledged that this linkage has been noted in other disciplines; and in fact, social motivation theory has stressed that these behaviour-related variables are indeed able to predict individual performance (Geen 1991). Work on role theories has also shown that those exhibiting particular roles drive team performance. For instance, Belbin (2002) observed that individuals in teams occupied eight distinct roles, and the *Shaper* role (those who are always keen on winning) and *Finisher* role (those who maintain a sense of urgency within the team) were focused heavily on task outcomes. Thus, the linkage between such attitudes and engagement is somewhat confirmatory and insightful for ISD. Furthermore, those studying behaviour patterns of IS practitioners have previously linked behaviours to performance. For instance, Gorla and Lam (2004) found high-performing programmers to be extroverted, and extroversion was observed to be a positive project management profile (Wang 2009) – particularly for effective communication. Here, however, we observed conscientious-related attitudes to be stronger indicators of ISD practitioners' engagement. Of course, our findings could also be interpreted from the inverse perspective. That is, as noted above re the communication of messages, perhaps Jazz developers' engagement in development work drove them to demonstrate achievement and insightful attitudes. Confirming such a linkage would be beneficial in terms of informing ISD team composition strategies, and such awareness may also help more generally with project diagnostics. We consider this issue further in the following section.

## IMPLICATIONS

Notwithstanding that we have examined a single context in this study, which may affect the work's generalizability (Runeson and Host 2009), we contend that the outcomes of this work are novel, and that these outcomes have implications for both research and ISD practice. From a research perspective, while previous work has observed the need for heterogeneity in team members' perspectives (Trimmer et al. 2002), this study provides tangible insights into how IS developers' attitudes are actually enacted during problem solving. Additionally, evidence provided in this work for the similarities and differences in attitudes expressed by developers reveals unique insights into ISD team dynamics. We observe that a given practitioner's social orientation is not a strong predictor of their communication volume; rather, their level of engagement in actual ISD tasks is more closely related to how frequently they communicate. These findings are somewhat contradictory to established thinking in the personality psychology space (Gorla and Lam 2004), and so provide an interesting and thought-provoking platform for replication studies. We found that those who were most conscientious were most engaged during the ISD process. This outcome also diverges somewhat from those studies that linked top performers to the extroversion personality trait (Wang 2009), and so would benefit from follow up work. In addition, our incremental application of multiple techniques in this work (comprising linguistic analysis, text mining and visualization, and statistical analysis) may inform those concerned with studying behavioural issues, particularly based on data drawn from repositories as opposed to being collected by the researcher in a live setting. Furthermore, should follow-up work provide a conclusive link between attitudes and engagement, such insights may inform requirements for decision support tools in aiding team composition (Licorish et al. 2009).

Beyond directions for follow up research work, our findings regarding team members' heterogeneity suggest that agile ISD project managers may benefit in terms of stable team climate if an approach to promote diversity among developers is utilized to inform team composition. In fact, while replication studies are encouraged to validate our findings in this work, the recommendation to promote diversity is well-established in the management and role theories domain. Further, we would assert that those ISD recruits who are outcome-driven are likely to provide the most value to their teams. Our findings suggest that ISD project managers should actively observe the attitudes of the cohort of developers under their leadership, as those who are outcome-driven are likely to play a critical role in their team's overall performance. Our evidence suggests that such individuals are likely to be central to their team's behavioural dynamics given that they also tend to communicate heavily. Their occupation of a central position should be encouraged as their outcome-driven attitudes may also propagate to the wider team (Licorish and MacDonell 2013a). That said, caution should also be exercised around the team's possible over-reliance on these achievement-driven members, as this may negatively affect the quality of the knowledge they are able to provide.

## THREATS TO VALIDITY

*Construct Validity:* The language constructs used to assess practitioners' attitudes in this study have been used previously to investigate this phenomenon (e.g., see Mairesse et al. (2007)). However, the adequacy of these constructs, and the suitability of the LIWC tool for studying IS practitioners' linguistic processes, may still be subject to debate. Communication was measured from messages sent around ISD tasks. Although project communication was enforced through the use of Jazz, these messages may not represent all of the developers' communication. Offsetting this concern is the fact that, as Jazz was developed as a globally distributed project, developers were required to use messages so that all other contributors (irrespective of their physical location) were aware of product and process decisions regarding each WI. Finally, we used task changes (including measures for tasks created, modified and resolved) to determine developers' engagements (Cataldo and Herbsleb 2008). However, all ISD tasks are not equal; some tasks may be more complex than others (e.g., a user experience task may not demand the same cognitive and mental rigor as that of a computational or coding-intensive feature). Nonetheless, such complexities would likely 'even out' over the entire project.

*Internal and External Validity:* We studied the artefacts of ISD practitioners from a single organization employing particular agile-like development practices. Work processes and the work culture at IBM Rational are likely to be specific to that organization and may not be representative of organization dynamics elsewhere, and particularly for environments that employ conventional waterfall processes (Boehm 2006). That said, Costa et al. (2011) confirmed that practitioners of the Jazz project exhibited similar coordination needs to practitioners of four projects operating in two distinct companies. Thus, we believe that our results may have some degree of applicability to similar, large-scale distributed projects conducted elsewhere.

## CONCLUSIONS

Motivated by a growing need to understand the human factors involved in ISD, along with evidence that ISD practitioners' communication artefacts can reveal integral details around what happens during the ISD process, we interrogated an instance of the IBM Rational Jazz repository to study the diversity of attitudes among IS developers, and the relationship between ISD practitioners' specific attitudes and their engagement. We used multiple analysis techniques, including linguistic analysis, text mining and visualization, and statistical analysis to show that ISD practitioners collectively expressed heterogeneous attitudes while working to deliver systems features, and there were some similarities across the expression of attitudes. Additionally, our findings revealed that those developers who had the most to say were also the most engaged during ISD, and these members were also most achievement-driven. This latter finding, although being supported by previous work on role theories, diverges from those that considered IS developers' personalities. We believe that, although drawn from a single case organisation, methodological insights and outcomes from this work provide an interesting platform for future research. Among our recommendations to those governing ISD projects, we advise diversity during team formation, and the active provision of support for those practitioners who are outcome-driven.

## ACKNOWLEDGEMENTS

We thank IBM for granting us access to the Jazz repository. S. Licorish carried out the research underpinning this paper while supported by an AUT University Vice-Chancellor's Doctoral Scholarship Award.